\documentclass[10 pt,prb,twocolumn,superscriptaddress,showpacs]{revtex4}

\usepackage[utf8]{inputenc}
\usepackage{graphics}
\usepackage{epsfig}
\usepackage{dcolumn}
\usepackage{bm}
\usepackage{amsmath}
\usepackage{amsfonts}
\usepackage{latexsym}
\usepackage{amssymb}
\usepackage{hyperref}
\allowdisplaybreaks[4]
\usepackage[normalem]{ulem}
\usepackage{color} 
\newcommand{\g}[1]{{\bf #1}}

\usepackage{subfigure}
\begin{document}

\title{Gutzwiller Wave-Function Solution for Anderson Lattice Model:\\ 
Emerging Universal Regimes of Heavy Quasiparticle States}

\author{Marcin M. Wysoki\'nski}
\email{marcin.wysokinski@uj.edu.pl}

\affiliation{Marian Smoluchowski Institute of Physics$,$ Jagiellonian University$,$ 
ulica \L{}ojasiewicza 11$,$ PL-30-348 Krak\'ow$,$ Poland}

\author{Jan Kaczmarczyk}
\email{jan.kaczmarczyk@ist.ac.at}

\affiliation{Marian Smoluchowski Institute of Physics$,$ Jagiellonian University$,$ 
ulica \L{}ojasiewicza 11$,$ PL-30-348 Krak\'ow$,$ Poland}

\affiliation{Institute of Science and Technology Austria$,$  Am Campus 1$,$ A-3400 Klosterneuburg$,$ Austria}

\author{Jozef Spa\l ek}
\email{ufspalek@if.uj.edu.pl}

\affiliation{Marian Smoluchowski Institute of Physics$,$ Jagiellonian University$,$ 
ulica \L{}ojasiewicza 11$,$ PL-30-348 Krak\'ow$,$ Poland}

\date{\today}

\begin{abstract}
 
The recently proposed diagrammatic expansion (DE) technique for the full Gutzwiller wave function (GWF)
is applied to the Anderson lattice model. This approach allows for a systematic evaluation of the expectation 
values with full Gutzwiller wave function in the finite dimensional systems. It introduces results extending in an essential manner 
those obtained by means of standard Gutzwiller approximation (GA) scheme which is 
variationally exact only in infinite dimensions.
Within the DE-GWF approach we discuss principal paramagnetic properties  
and their relevance to the heavy fermion systems. 
We demonstrate the formation of an effective, narrow $f$-band 
originating from atomic $f$-electron states and 
subsequently interpret this behavior as a {\it direct itineracy} of $f$-electrons; 
it represents  a combined 
effect of both the hybridization and the correlations reduced by the 
Coulomb repulsive interaction. 
Such feature is absent on the level of GA 
which is equivalent to the zeroth order of our expansion.
Formation of the hybridization- and electron-concentration-dependent narrow 
$f$-band rationalizes common assumption of such dispersion of $f$ levels in the phenomenological modeling 
of the band structure of CeCoIn$_5$.
Moreover, it is shown that the emerging $f$-electron {\it direct itineracy} leads in a natural manner to three
physically distinct regimes within a single model, that are frequently discussed for 4$f$- or 5$f$- electron
compounds as separate model situations. 
We identify these regimes as: (i) mixed-valence regime, (ii) Kondo-insulator border regime,
and (iii) Kondo-lattice limit when the $f$-electron occupancy 
is very close to the $f$-states half-filling, $\langle\hat n_{f}\rangle\rightarrow1$.
The nonstandard features of emerging correlated quantum liquid state are stressed.

\end{abstract}


\maketitle
\section{Introduction and Motivation}\label{I}

Heavy fermion systems (HFS) belong to the class of quantum materials with strongly correlated 
4$f$ or 5$f$ electrons. They exhibit unique properties resulting from their universal electronic features  
(e.g. very high density of states at the Fermi level)
almost independent of their crystal structure. Among those unique 
properties are: (i) enormous effective masses in the Fermi-liquid state, as demonstrated through 
the linear specific heat coefficient \cite{Ott1975,Stewart1984, Steglich1991,Fulde1988,Ott1987} and their 
direct spin-dependence in the de Haas-van Alphen measurements \cite{Citro1999,Sheikin2003,McCollam2005}, 
(ii) Kondo-type screening of localized or almost localized $f$-electron magnetic moments by the 
conduction electrons \cite{Doradzinski1997,Doradzinski1998}, 
(iii) unconventional superconductivity, appearing frequently at the border or coexisting with magnetism \cite{Pfleiderer2009}, 
and (iv) abundance of quantum critical points and associated with them non-Fermi
(non-Landau) liquid behavior \cite{Stewart2001,Wolfle2007,Lonzarich2005}.

The Anderson lattice model (ALM), also frequently referred to as
periodic Anderson model, and its derivatives: the Kondo\cite{Lacroix1979,Hewson1993,Auerbach1987} 
and the Anderson-Kondo\cite{Howczak2012,Howczak2013} 
lattice models, capture the essential physics of HFS.
 Although, the class of exact solutions is known for this model\cite{g1,g2,g3,g4}, 
they are restricted in the parameter space. Thus, for thorough investigation of the model 
properties the approximate methods are needed.   
One of the earliest theoretical approaches for 
 the models  with a strong Coulomb repulsion was the 
variational Gutzwiller wave function (GWF) method \cite{Varma1986,strack1993,Rice1985,Rice1986, Miyake1986, Fazekas1999}. 
However, despite its simple and physically transparent form, a direct analytic evaluation of 
the expectation values with full GWF cannot be carried out rigorously 
for arbitrary dimension and spatially unbound systems. 

\vspace{-0.1cm}
\begin{center}
  \begin{figure}[t]
   \includegraphics[width=0.5\textwidth]{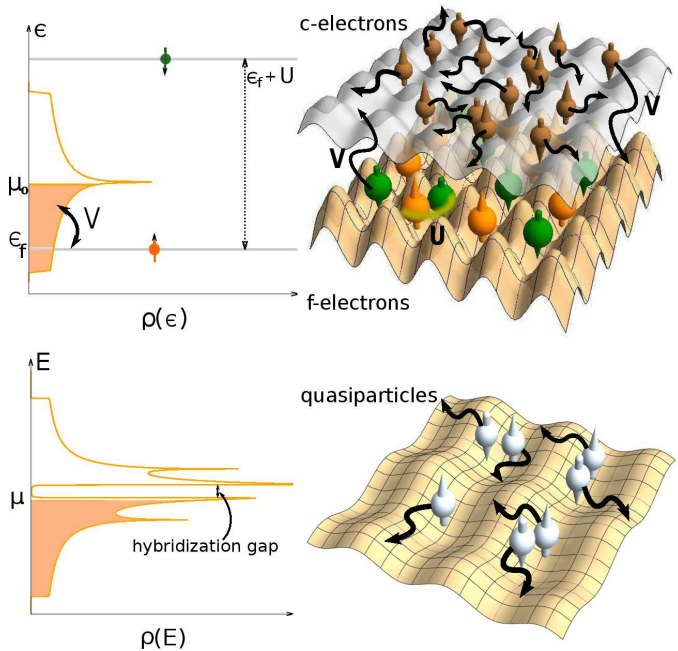}
   \caption{(Color online) Upper part: Schematic representation of the two-orbital Anderson lattice model
   with initially localized $f$- and delocalized $c$-electrons, 
   and hybridization between them. Bottom part: Emerging quasiparticle
   states in the hybridized bands of correlated particles. On the left: the shapes of the density 
   of states in the respective situations.}   \label{Fig1}
  \end{figure}
 \end{center}
 \vspace{-0.8cm}

One of the ways of overcoming this difficulty is the so-called 
Gutzwiller Approximation (GA), in which only local two-particle 
correlations are taken into account when evaluating the expectation values.
GA provides already a substantial insight into 
the overall properties of strongly
correlated systems \cite{Rice1985,Kotliar1986,Spalek1987,Gebhard1991,
Dorin1992,Dorin1993,Doradzinski1997,Doradzinski1998,Bunemann1998}.
Moreover, this approach has been reformulated recently to the so-called 
{\it statistically-consistent Gutzwiller approximation} (SGA) scheme 
and successfully applied to a number of problems
involving correlated electron systems
\cite{Jedrak2011,kaczmarczyk2011,Howczak2013,
Abram2013,Kadzielawa2013,Zegrodnik2014,Wysokinski2014,WysokinskiAbram2014, WysokinskiAbram2015}.
Among those, a concrete application has been a
microscopic description of the fairly complete magnetic phase diagram 
of UGe$_2$ \cite{WysokinskiAbram2014, WysokinskiAbram2015} which provided quantitatively
correct results, even without 
taking into account the 5$f$-orbital degeneracy due to uranium atoms.

An advanced method of evaluating the
expectation values for GWF 
is the variational Monte Carlo technique (VMC) 
\cite{Edegger2006,Lugas2006,Raczkowski2007,Edegger2007,Chou2012,Schmalian,Watanabe,Fabrizio2013,Watanabe2015}.
However, this method is computationally expensive and suffers from the 
system-size limitations. Though, one must note that the VMC method allows for extension of GWF
by including e.g. Jastrow intersite factors \cite{Jastrow}.

Here we use an alternative method of evaluating the expectation values
for GWF, namely a systematic diagrammatic expansion for the 
Gutzwiller wave function (DE-GWF) \cite{buenemann2012,Jan_hubbard,Jan_tj,PhilMag_Jan,pss}. 
This method was formulated initially for the Hubbard model in two dimensions in the context 
of Pomeranchuk instability\cite{buenemann2012}, and applied subsequently to 
the description of high-temperature
superconductivity for the Hubbard\cite{Jan_hubbard,pss} and the $t$-$J$\cite{Jan_tj} models.
In the zeroth order of the expansion this approach straightforwardly reduces 
to the GA\cite{Jan_tj}. For the one-dimensional Hubbard model it converges 
\cite{buenemann2012} to the exact GWF results. Within DE-GWF a larger variational space 
can be sampled than within the alternative VMC technique because the long-range components of the effective
Hamiltonian are accounted for naturally. The DE-GWF method (truncated to match the variational spaces) 
reproduces the results of VMC with  improved accuracy (as shown for the $t$-$J$ \cite{Jan_tj} and the Hubbard models 
\cite{Jan_hubbard}).
Additionally, the method works also in the thermodynamic limit.
In effect, the approach is  well suited to capture   
subtle effects, e.g. those related to the topology of the Fermi surface
in the correlated state \cite{buenemann2012} or  
the investigated here formation of a narrow $f$-electron band.

In this study, we extend the DE-GWF approach to discuss 
principal paramagnetic properties within ALM.
The emergence of the quasiparticle picture  
is schematically illustrated in Fig.\ref{Fig1}. 
Explicitly, we investigate the shape of the quasiparticle density of states (DOS, $\rho(E)$),
evolving with the increasing order of the expansion, $k$.
For $k>0$  the hybridization gap widens up 
with respect to that in GA ($k=0$ case) and DOS peaks are
significantly pronounced. Moreover, we investigate
the DOS at the Fermi level ($\rho(E_F)$) evolution with the increasing the hybridization
strength $|V|$ -- total electron concentration $n$, plane, as it
is a direct measure of the heavy-quasiparticle effective mass. 
We find that this parameter is significantly
enhanced for $k>0$, mainly in the low hybridization limit 
and at the border of the Kondo-insulating state.
Furthermore, we trace the contribution coming from the originally  
localized $f$-electrons (cf. Fig. \ref{Fig1} - upper part) 
to the quasiparticle spectrum with
the increasing order of the expansion. For $k>0$, 
$f$-quasiparticles effectively acquire a nonzero bandwidth 
(up to 6\% of the conduction bandwidth) as a combined effect of 
both interelectronic correlations and hybridization.  
 
Assumption of a narrow $f$ band existence has recently been made
in a phenomenological modeling of the heavy fermion compound 
CeCoIn$_5$ band structure \cite{Yazdani2012,allan2013,dyke2014}.
We show that the emergence of such a band, 
absent in GA ($k=0$), is an evidence
of the $f$-electron direct itineracy explained later.
To quantify this itineracy we introduce the parameter $w_f$ -
the width of the effective, narrow $f$-band. 
On the hybridization
strength -- total electron concentration, $|V|$ -- $n$ plane, $w_f$ is 
significantly enlarged in the three distinct regimes,
which we identify respectively as the mixed-valence, Kondo/almost Kondo insulating, 
and the Kondo-lattice regimes 
(when $f$-electron concentration is close to the half-filling, 
i.e., when $\langle\hat n_{f}\rangle\rightarrow1$). 
These physically distinct regimes are frequently discussed and identified in various experiments 
\cite{Holmes2004,Slebarski2005,Pfleiderer2009,Stewart1984,Slebarski2010,Kaczorowski2009,Kaczorowski2010}
and in theory \cite{Watanabe2010,Howczak2013,Dorin1992,Ueda1997}.

The structure of the paper is as follows.  In Sec. \ref{sec2} we describe the ALM Hamiltonian 
and define the Gutzwiller variational wave function in a nonstandard manner. 
In Sec. \ref{sec2b} we derive the
DE-GWF method for ALM and determine the effective single-particle two-band Hamiltonian. 
In Sec. \ref{sec3}
we present results concerning paramagnetic properties: the quasiparticle spectrum, 
the resultant density of states at the Fermi level, and formation of an effective narrow 
$f$-electron band out of initially
localized states. In Appendix \ref{B} we discuss the equivalence of the zeroth-order DE-GWF approach with GA. 
In Appendix \ref{A} we present some technical details of DE-GWF technique.
 
  \section{Model Hamiltonian and Gutzwiller wave function} \label{sec2}

Our starting point is the Anderson lattice model (ALM)  with the
chemical potential $\mu$ and expressed through Hamiltonian
\begin{equation}
\begin{gathered}
 \mathcal{\hat H}={\sum_{\g i,\g j,\sigma}} t_{\g i\g j}\hat c_{\g i\sigma}^\dagger\hat c_{\g j\sigma}
-\sum_{\g i,\sigma}\mu\hat n^c_{\g i\sigma}
 +\sum_{\g i,\sigma}(\epsilon_f-\mu)\hat n^f_{\g i\sigma}\\+ 
U\sum_\g i \hat n^f_{\g i\uparrow} \hat n^f_{\g i\downarrow}+
\sum_{\g i,\g j,\sigma}(V_{\g i\g j}\hat f_{\g i\sigma}^\dagger
\hat c_{\g j\sigma}+V^*_{\g i\g j}\hat c_{\g i\sigma}^\dagger\hat f_{\g j\sigma}),
\label{alm}
\end{gathered}
\end{equation}
where $\g i=(i_x,i_y)$ (and similarly $\g j$) is the two-dimensional site 
index, $\hat f_{\g{i}\sigma}$ ($\hat f^\dagger_{\g{i}\sigma}$) and $\hat c_{\g{i}\sigma}$ ($\hat c^\dagger_{\g{i}\sigma}$) 
are the annihilation (creation) operators related to 
$f$- and $c$- orbitals respectively, and $\sigma=\uparrow,\downarrow$ is the $z$-component 
direction of the spin. 
We assume that the hopping   in the conduction 
band takes place only between the nearest neighboring 
sites, $t_{\g i\g j}\equiv t\delta_{|\g i-\g j|,1}$, 
 the hybridization has the simplest onsite character \cite{nota}, $V_{\g i\g j}=V\delta_{\g i,\g j}$, the local 
 Coulomb repulsion on the $f$ orbital has the amplitude $U$,
 and the initially atomic $f$ states are located at the energy $\epsilon_f$.  
In the following $|t|$ is used as the energy unit. 

 Gutzwiller wave function (GWF) is constructed from the uncorrelated Slater determinant $| \psi_0 \rangle$ 
by projecting out fraction of the local double $f$-occupancies  
by means of the Gutzwiller projection operator $\hat P_G$,
\begin{equation}
 | \psi_G \rangle\equiv\hat P_G|\psi_0 \rangle\equiv \prod_{\g i}\hat P_{G;\g i}| \psi_0 \rangle.
 \label{psig}
\end{equation}
In the GA approach when only a single $f$ orbital (in the present case) is correlated the projection operator can be defined by 
\begin{equation}
\hat P_{G;\g i}\equiv 1-(1-g)\hat n_{\g i\uparrow}^f\hat n_{\g i\downarrow}^f, \label{std}
\end{equation}
where $g$ is a variational parameter. Such form allows for interpolating between the 
fully correlated ($g=0$) and the uncorrelated ($g=1$) limits. Equivalently  one can consider average number of
doubly occupied states, $\langle \hat n_{\g i\uparrow}^f \hat n_{\g i\downarrow}^f \rangle  \equiv d^2$ as 
a variational parameter. 

The Gutzwiller projection operator can be selected differently as 
proposed in Ref. \onlinecite{Gebhard1990}, namely
\begin{equation}
\hat P_{G;\g i}^\dagger\hat P_{G;\g i}\equiv \hat P_{G;\g i}^2={\bf1} +x\hat d_{\g i}^{HF}.
\label{PI}
\end{equation}
In the above relation $x$ is a variational parameter and for the paramagnetic and 
translationally invariant system we define Hartree-Fock (HF) operators of the form
\begin{equation}
\hat d_{\g i}^{HF}\equiv\hat n^{HF}_{\g i\uparrow}\hat n^{HF}_{\g i\downarrow}=
(\hat n_{\g i\uparrow}^f -n_{0f})(\hat n_{\g i\downarrow}^f -n_{0f}),
\end{equation}
where $n_{0f}$ denotes average occupation of a single $f$ state and spin $\sigma$ in the 
uncorrelated state, $|\psi_0\rangle$, i.e.,
${n_{0f}=\langle\hat f_{\g{i}\sigma}^\dagger\hat f_{\g{i}\sigma}\rangle_0}$.
Hereafter the shortened notation for the expectation values is used, i.e., 
$\langle\psi_0|...|\psi_0\rangle\equiv\langle...\rangle_0$.
Strictly speaking, although, $\hat d_{\g i}^{HF}$ has not the Hartree-Fock
form of the double occupancy operator, the HF superscript has its meaning as 
the property, $\langle\hat d_{\g i}^{HF}\rangle_0\equiv0$ is preserved.

On the other hand, the Gutzwiller projection operator can be defined in general form as
\begin{equation}
 \hat P_{G;\g i}=\sum_\Gamma \lambda_\Gamma\mid\Gamma\rangle_{\g i}\langle\Gamma\mid_{\g i},
\end{equation}
with variational parameters 
$\lambda_\Gamma \in \{\lambda_0,\lambda_\uparrow,\lambda_\downarrow,\lambda_d\}$ that 
characterize the possible occupation probabilities for the four possible atomic Fock $f$-states 
${\mid\Gamma\rangle_{\g i}\in\{\mid 0\rangle_{\g i},\mid\uparrow\rangle_{\g i},
\mid\downarrow\rangle_{\g i},\mid\uparrow\downarrow\rangle_{\g i}\}}$.

Relation (\ref{PI}) couples $\lambda_\Gamma$ and $x$,
reducing the number of independent variational parameters  to one. Explicitly,
we may express the parameters $\lambda_\Gamma$ by the coefficient $x$, 
\begin{equation}
 \begin{split}
 \lambda_0^2&=1+xn_{0f}^2,\\
\lambda_\sigma^2=\lambda_{\bar \sigma}^2\equiv\lambda_s^2&=1-xn_{0f}(1-n_{0f}),\\
\lambda_d^2&=1+x(1-n_{0f})^2. \label{dd}
\end{split}
\end{equation}
As the parameters $\lambda_\Gamma$ and $x$ are coupled by the conditions (\ref{dd}), 
there is a freedom of choice of the variational parameter; in this work we have selected $x$. 
The parameter $x$ covers the same variational space as $g$ in GA. Additionally,
the projector (\ref{PI}) leads to much faster convergence than (\ref{std}) (cf. Ref. \onlinecite{buenemann2012}).
From (\ref{PI}) it is clear that $x=0$ corresponds to the uncorrelated limit. The other 
extremity, the fully correlated state is reached for $x=\max\{x(\lambda_d=0),x(\lambda_0=0)\}$.
This leads to the bounds $\max\{\frac{-1}{(1-n_{0f})^2},\frac{-1}{(n_{0f})^2} \}\leq x\leq0$.
The minimal value is $x=-4$ for $n_{0f}=0.5$.

The method is suitable
for an arbitrary filling of the $f$ orbital. However, due to the fact that present work
is mainly addressed to the description of the Ce-based compounds, we study the regime in which the
$f$-orbital filling either does not exceed unity or is only slightly larger. 
Precisely, in the all figures presented here the $f$-orbital filling is never larger than 1.05.

 \section{DE-GWF Method} \label{sec2b}
 \subsection{General scheme}\label{gsc}
 In this section we present general implementation of the DE-GWF method.
The procedure is composed of the following steps:
\begin{enumerate}
\item Choice of initial state $|\psi_0\rangle$.
 \item Evaluation of $\langle\mathcal{\hat H}\rangle_G\equiv\frac{\langle\psi_G\mid\mathcal{\hat H}\mid\psi_G\rangle}{\langle\psi_G\mid\psi_G\rangle}$ 
 for selected $|\psi_0\rangle$ - cf. Sec. \ref{b2sec}.
 \item Minimization of $\langle\mathcal{\hat H}\rangle_G$ with respect to the variational parameter (here $x$).
 \item Construction of the effective single particle Hamiltonian determined by 
 $\delta\mathcal{\hat H}^{\rm eff}(|\psi_0\rangle)=\delta\mathcal{\hat H}(|\psi_0\rangle)$ - cf. Sec. \ref{c2sec}.
 \item Determination of $|\psi'_0\rangle$ as a ground state of the effective Hamiltonian - cf. Sec. \ref{d2sec}.
 \item Execution of the self-consistent loop: starting again from the step 1 with  $|\psi'_0\rangle$ until a satisfactory convergence, 
 i.e., $|\psi'_0\rangle=|\psi_0\rangle$, is reached.
\end{enumerate}

Steps 4 and 5 ensure that the final form of $|\psi_0\rangle$ represents the optimal 
choice which minimizes the ground state energy $\langle\mathcal{\hat H}\rangle_G$.
The DE-GWF method with respect to other related methods, GA and VMC, introduces a new technique for
evaluating the expectation value of the correlated Hamiltonian with GWF (step 2 
of the above procedure). 
In particular, it provides an important improvement as, e.g., for GA only single sites 
in the lattice contain the projection whereas the remaining environment does not.
GA leads e.g. to the inability of obtaining the superconducting phase in the Hubbard model \cite{Jan_hubbard}. 
On the other hand, the VMC method tackles that  problem properly, but 
at the price of extremely large computing power needed. This leads to
the lattice size limitations (typically up to 20x20 sites) 
and a limited distance of real space intersite correlations taken into account.

In this respect, DE-GWF introduces,
in successive orders of the expansion, correlations to the
environment of individual sites (beyond GA), as well as converges in a 
systematic manner to the full GWF solution. Also,
DE-GWF was shown to provide results of better accuracy than VMC \cite{Jan_tj},
and additionally, is free from the finite-size limitations.
It also demands definitely less computational power than VMC.
Thus in general, this method is capable of treating more complex
problems with GWF. 
On the other hand, DE-GWF is tailored specifically for GWF, while VMC 
allows for starting from different forms of variational wave function e.g., adding 
the Jastrow factors\cite{Jastrow,Watanabe2015}.

 \subsection{Diagrammatic expansion}\label{b2sec}
The key point of the variational procedure is the calculation of the 
expectation value of Hamiltonian (\ref{alm}) 
with GWF $|\psi_G\rangle$ (point 1 from the scheme in Sec. \ref{gsc}), by starting from the expression
\begin{equation}
\begin{gathered}
 \langle\mathcal{\hat H}\rangle_G\equiv\frac{\langle\psi_G\mid\mathcal{\hat H}\mid\psi_G\rangle}
 {\langle\psi_G\mid\psi_G\rangle}=\frac{\langle\psi_0\mid \hat P_{G}\mathcal{\hat H}\hat P_{G}\mid\psi_0\rangle}
 {\langle\psi_0\mid\hat P_{G}^2\mid\psi_0\rangle}.\label{eq8}
\end{gathered}
 \end{equation}

We use the DE-GWF technique \cite{buenemann2012,Jan_hubbard,Jan_tj,PhilMag_Jan}, 
based on the expansion of the expectation values appearing in Eq. (\ref{eq8}) 
in the power series in variational parameter $x$, with the highest power 
representing number of correlated vertices assumed to be correlated in the environment - besides local ones. 
This method is systematic in the sense that the zeroth order corresponds to GA \cite{Edegger2007}, 
whereas with the increasing order the full GWF solution is approached. 
Explicitly, we determine expectation values with respect to GWF of any product operator 
originating from the starting Hamiltonian~(\ref{alm})
${{\mathcal{\hat O}_{\g i(\g j)}}=\{\hat c_{\g i\sigma}^\dagger\hat c_{\g j\sigma},
 \hat n^c_{\g i\sigma},\hat n^f_{\g i\sigma},\hat n^f_{\g i\uparrow} \hat n^f_{\g i\downarrow},
 \hat f_{\g i\sigma}^\dagger\hat c_{\g j\sigma},\hat c_{\g i\sigma}^\dagger\hat f_{\g j\sigma}\}}$. 
 This is executed by first accounting for the projection part on the site $\g i(\g j)$ - {\it external} vertices
 (e.g., computing ${{\hat{\mathcal{O}}^G_{\g i(\g j)}}\equiv 
\hat P_{G;\g i}(\hat P_{G;\g j})\hat{\mathcal{O}}_{\g i(\g j)}(\hat P_{G;\g j})\hat P_{G;\g i}}$, see below)
and then, including one-by-one correlations (terms) to the other sites $\g l\neq\g i,\g j$ - {\it internal} vertices.

Formally, the procedure starts in effective power expansion in $x$ 
of all relevant expectation values
\begin{equation}
\begin{gathered}
 \langle\psi_G\mid \hat{\mathcal{O}}_{\g i (\g j)} \mid\psi_G\rangle=
 \Big\langle\hat{{\mathcal{O}}}^G_{\g i(\g j)}\prod_{\g l\neq\g i,\g j} \hat P_{G;l}^2 \Big\rangle_0\\
 =\sum_{k=0}^\infty\frac{x^k}{k!}{\sum_{\g l_1,...,\g l_k}}'
 \langle\hat{{\mathcal{O}}}^G_{\g i(\g j)}\hat d^{HF}_{\g l_1,...,\g l_k}\rangle_0,
 \label{expe}
\end{gathered}
 \end{equation}
where  
$\hat d^{HF}_{\g l_1,...,\g l_k}\equiv\hat d^{HF}_{\g l_1}\cdots\hat d^{HF}_{\g l_k}$.
The prime in the multiple summation denotes restrictions:
$\g l_p\neq\g l_{p'}$, and $\g l_p\neq\g i,\g j$ for all $p,p'$. $k$ is the order of the expansion.
Note that for $k=0$ we obtain $\langle\psi_G| \hat{\mathcal{O}}_{\g i (\g j)}|\psi_G\rangle=
\langle\hat{{\mathcal{O}}}^G_{\g i(\g j)}\rangle_0$. This means that 
the projection operators act only locally (i.e., only the sites $\g i$ 
and $\g j$ are affected) and in this case we recover the GA results (for a details discussion of the 
equivalence see Appendix \ref{B}).
Expectation values in (\ref{expe}) can now be calculated by means of the Wick's theorem
in its real-space version,
as they involve only products averaged with $|\psi_0\rangle$. 
Such power expansion in $x$ allows for taking into account long-range 
correlations between $k$ {\it internal} sites ($\g l_1,...,\g l_k$) and the {\it external} ones ($\g i,\g j$).
It must be noted that it is not a perturbative 
expansion with respect to the small parameter $x$. Instead,
the expansion should be understood as an analytic series with the order 
determined by  the number of correlated {\it internal vertices} 
taken in the nonlocal environment. For $k=\infty$, the full GWF solution would be obtained.
However, on the basis of our results, a satisfactory results for the expansion 
in ALM case are reached already starting from $k=3$. 

As said above, the expectation values $\langle...\rangle_0$ in Eq.~(\ref{expe}) can be evaluated 
by means of the Wick's theorem. Then, the terms with $k$ {\it internal} sites can be visualized as 
diagrams with $k$ internal and $1$ (or $2$) {\it external} vertices. 
The lines connecting those vertices are defined as,
\begin{equation}
\begin{split}
C_{\g{i}\g{j}}&\equiv \langle \hat c_{\g{i}\sigma}^\dagger\hat c_{\g{j}\sigma} \rangle_0,\\
W_{\g{i}\g{j}}&\equiv \langle \hat f_{\g{i}\sigma}^\dagger\hat c_{\g{j}\sigma} \rangle_0,\\
F_{\g{i}\g{j}}&\equiv \langle \hat f_{\g{i}\sigma}^\dagger\hat f_{\g{j}\sigma} \rangle_0-\delta_{\g{i}\g{j}}n_{0f}.
\label{lines1}
\end{split}
\end{equation}
By constructing the projector operator (\ref{PI}), we have eliminated all the diagrams 
with the local $f$-orbital  contractions ($\langle \hat f_{\g{i}\sigma}^\dagger\hat f_{\g{i}\sigma} \rangle_0$), 
the so-called {\it Hartree bubbles}.
This procedure, as discussed explicitly in Ref.~\onlinecite{buenemann2012}, leads to significantly
faster convergence than that with the usual Gutzwiller projector, with the variational parameter $g$ \cite{g5}. 
It constitutes the main reason for the efficiency of the DE-GWF method.
Finally, all the expectation values with respect to GWF are normalized by  
$\langle\psi_G|\psi_G\rangle$ (cf.~Eq.~(\ref{eq8})).
However, through the linked-cluster theorem \cite{Fetter}, the terms coming from expansion of $\langle\psi_G|\psi_G\rangle\equiv
\langle\psi_0|\hat P_{G}^2|\psi_0\rangle$ cancel out with  all disconnected diagrams 
appearing in the numerator of Eq.~(\ref{eq8}).
In effect, the expectation values can be expressed in the closed form by the 
diagrammatic sums ${S\in\{T^{cc(1,1)}_{\g{ij}}, T^{fc(1,1)}, T^{fc(3,1)},I^{c(2)}_{}, I^{f(2)}_{},I^{f(4)}\}}$,
defined in Appendix \ref{A}, what leads to the following resultant expression for the ground state energy:
\begin{equation}
\begin{gathered}
 \frac{\langle\mathcal{\hat H}\rangle_G}{ L }
 =\frac{2}{ L }{\sum_{\g i,\g j}}t_{\g i\g j}{T}^{cc(1,1)}_{\g i\g j}-2 \mu I^{c(2)}\\
+2(\epsilon_f-\mu)\left(n_{0f}+(1+xm) I^{f(2)}+\gamma I^{f(4)}\right)\\
+U\lambda_d^2\left(d_0+2n_{0f}I^{f(2)}+(1-xd_0)I^{f(4)}\right)\\
+4 V\left(\alpha T^{fc(1,1)} +
\beta T^{fc(3,1)} \right),
\label{Ham}
\end{gathered}
\end{equation}
where the trivial sums $\sum_\sigma=2$ and $\sum_{\g i}= L $  have already been included.
Parameters $\{\alpha,\beta,\gamma,m,d_0\}$ are all functions of $n_{0f}$ and $x$ (cf. Appendix \ref{A}, Eq. (\ref{A4})).
For $k=0$, only the diagrammatic sums ${T}^{cc(1,1)}_{\g i\g j},I^{c(2)}$ and $T^{fc(1,1)}$ do not vanish and we reproduce
the standard GA result; the Coulomb energy reduces to $U\lambda_d^2d_0$ and hybridization to 
$4V\alpha\langle\hat f_{\g i}^\dagger\hat c_{\g i}\rangle_0$, whereas the diagrammatic sums for $c$-band only are trivial
(cf. the discussion in Appendix \ref{B}).

The expectation value $\langle\mathcal{\hat H}\rangle_G$ calculated diagrammatically is minimized next with respct
to the variational parameter $x$ (step 3 in the scheme in Sec. \ref{gsc}).

\subsection{Effective quasiparticle Hamiltonian}\label{c2sec}

The next step in our procedure (step 4 in the scheme in Sec. \ref{gsc}) is the mapping of the correlations contained in  
$\langle\psi_G|\mathcal{\hat H}|\psi_G\rangle/\langle\psi_G|\psi_G\rangle$ 
onto the corresponding uncorrelated expectation value $\langle\psi_0|\mathcal{\hat H}^{\rm eff}|\psi_0\rangle$.
It is realized via the condition that the minima of the expectation values of both Hamiltonians 
coincide for the same equilibrium values of lines (\ref{lines1})  and $n_{0f}$, which define $|\psi_0\rangle$. 
Note that the present formulation of this step of our minimization procedure is equivalent to those previously 
used \cite{buenemann2012,Jan_hubbard,Jan_tj,PhilMag_Jan,pss}. Explicitly,
\begin{equation}
\begin{gathered}
\delta\langle \mathcal{\hat H}^{\rm eff}\rangle_0 (C,F,W,n_{0f})= \delta\langle \mathcal{\hat H}\rangle_G (C,F,W,n_{0f})\\
=\frac{\partial \langle \mathcal{\hat H}\rangle_G}{\partial C}\delta C 
 +\frac{\partial \langle \mathcal{\hat H}\rangle_G}{\partial W}\delta W
 +\frac{\partial \langle \mathcal{\hat H}\rangle_G}{\partial F}\delta F
 +\frac{\partial \langle \mathcal{\hat H}\rangle_G}{\partial n_{0f}}\delta n_{0f},
\end{gathered}
\end{equation}
where skipping lattice indices for lines means that we consider each of them separately.
It leads directly to the following form of the effective single-particle two-band Hamiltonian with non-local
interband hybridization, i.e, 
\begin{equation}
\begin{gathered}
 \mathcal{\hat H}^{\rm eff}=\sum_{\g i,\g j,\sigma}t_{\g i \g j}^{c}\hat c_{\g i\sigma}^\dagger\hat c_{\g j\sigma}+
\sum_{\g i,\g j,\sigma}t_{\g i \g j}^{f}\hat f_{\g i\sigma}^\dagger\hat f_{\g j\sigma}\\
+\sum_{\g i,\g j,\sigma}(V_{\g i\g j}^{fc}\hat c_{\g i\sigma}^\dagger\hat f_{\g j\sigma}+{\rm H.c.}),\label{effH}
\end{gathered}
\end{equation}
where the effective hopping and hybridization parameters are derivatives with respect to lines, 
\begin{equation}
 \begin{split}
  t_{\g i\g j}^{c}&=\frac{\partial  \langle \mathcal{\hat H}\rangle_G}{\partial C_{\g i\g j}},\ \ \ \
  V_{\g i\g j}^{fc}=\frac{\partial  \langle \mathcal{\hat H}\rangle_G}{\partial W_{\g i\g j}},\\
 t_{\g i\g j}^{f}&=\frac{\partial  \langle \mathcal{\hat H}\rangle_G}{\partial F_{\g i\g j}},\ \ \ \ \ \
 t_{\g i\g i}^{f}=\frac{\partial  \langle \mathcal{\hat H}\rangle_G}{\partial n_{0f}}.
\label{te}
\end{split}
\end{equation}

\subsection{Determination of $|\psi'_0\rangle$ }\label{d2sec}
In this section we determine $|\psi'_0\rangle$ as a ground state of $\mathcal{\hat H}^{\rm eff}$ (point 5 from the scheme in Sec. \ref{gsc}).

In order to obtain the effective dispersion relations for $c$- and $f$-electrons and the $\g k$-dependent 
hybridization we use the lattice Fourier transform
\begin{equation}
\begin{gathered}
 \epsilon_{\g k}^{c(f)}=\frac{1}{ L } \sum_{\g{i,j}}e^{i(\g{i}-\g{j})\g k}t_{\g i\g j}^{c(f)},\\ 
V_{\g{k}}^{cf}=\frac{1}{ L } \sum_{\g{i,j}}e^{i(\g{i}-\g{j})\g k}V_{\g i \g j}^{fc}.
\end{gathered}
\end{equation}
In this manner, we reduce the many-body problem to the effective single-quasiparticle picture (cf. Fig. \ref{Fig1})
described by the effective two-band Hamiltonian. The 2x2 -matrix representation of Eq. (\ref{effH}) resulting from such
a transform, has the following form 
\begin{equation}
\begin{split}
 \mathcal{\hat H}^{\rm eff}&=\sum_{\g{k},\sigma}  (\hat c_{\g{k}\sigma}^\dagger\ \hat f_{\g{k}\sigma}^\dagger)
\begin{pmatrix}
 \epsilon_{\g{k}}^{c}&V^{cf}_{\g{k}}\\
V^{cf}_{\g{k}}& \epsilon_{\g{k}}^{f} \\ 
\end{pmatrix}
\begin{pmatrix}
\hat c_{\g{k}\sigma}\\
\hat f_{\g{k}\sigma}
\end{pmatrix}\\
&= 
\sum_{\g{k},\sigma}  (\hat c_{\g{k}\sigma}^\dagger\ \hat f_{\g{k}\sigma}^\dagger)
\mathcal{T}^\dagger
 \begin{pmatrix}
 E_{\g{k}+}&0\\
0&E_{\g{k}-} \\ 
\end{pmatrix}
\mathcal{T}
\begin{pmatrix}
\hat c_{\g{k}\sigma}\\
\hat f_{\g{k}\sigma}
\end{pmatrix},\label{he}
\end{split}
\end{equation}
where the eigenvalues, $E_{\g{k}\pm}$ of the above Hamiltonian are
\begin{equation}
 E_{\g{k}a}=\xi_{\g{k}}^+
+ a \sqrt{(\xi_{\g{k}}^-) ^2+(V_{\g{k}}^{cf})^2},\label{eigen}
\end{equation}
where $a\equiv\pm1$ differentiates between the two hybridized bands. For convenience, we have defined
\begin{equation}
 \xi_{\g{k}}^+\equiv\frac{\epsilon_{\g{k} }^c +\epsilon_{\g{k} }^f}{2}\ \ \ {\rm and}  \ \ \xi_{\g{k}}^-\equiv\frac{\epsilon_{\g{k} }^c -\epsilon_{\g{k} }^f}{2}.
\end{equation}
$\mathcal{T}$ in Eq. (\ref{he}) is the unitary transformation matrix to the basis in which   
 $ \mathcal{\hat H}^{\rm eff}$ is diagonal, defined as
\begin{equation}
\mathcal{T}=  
\begin{pmatrix}
 u_+&u_-\\
u_-&-u_+ \\ 
\end{pmatrix},
\end{equation}
where
\begin{equation}
 \begin{gathered}
  u_\pm=\sqrt{\frac{1}{2}\left(1\pm\frac{\xi_{\g{k}}^-}{\sqrt{(\xi_{\g{k}}^-)^2+(V_{\g{k}}^{cf})^2}}\right)}.
  \end{gathered}
\end{equation}
It is now straightforward to obtain the principal correlation functions (lines), i.e.
\begin{equation}
 \begin{split}
\langle \hat c_{\g{k}\sigma}^\dagger\hat c_{\g{k}\sigma} \rangle_0&=u_+^2\Theta(E_{{\g{k}} +})+u_-^2\Theta(E_{{\g{k}} -}),\\
\langle \hat f_{\g{k}\sigma}^\dagger\hat c_{\g{k}\sigma} \rangle_0&=u_+u_-\big(\Theta(E_{{\g{k}} +})-\Theta(E_{{\g{k}} -})\big),\\
\langle \hat f_{\g{k}\sigma}^\dagger\hat f_{\g{k}\sigma} \rangle_0&=u_-^2\Theta(E_{{\g{k}} +})+u_+^2\Theta(E_{{\g{k}} -}),  
 \end{split}
\end{equation}
where $\Theta(E)$ denotes the Heaviside step function and plays the role of 
an energy cutoff for respective quasiparticle bands energies (\ref{eigen}).
Using the reverse Fourier transformation we obtain self-consistent equations for lines and $n_{0f}$,
\begin{equation}
 \begin{split}
  C_{\g{i}\g{j}}&=\frac{1}{ L }\sum_{\g{k}a}\langle \hat c_{\g{k}\sigma}^\dagger\hat c_{\g{k}\sigma} \rangle_0\ e^{i(\g i- \g j)\g{k}},\\
  W_{\g{i}\g{j}}&=\frac{1}{ L }\sum_{\g{k}a}\langle \hat f_{\g{k}\sigma}^\dagger\hat c_{\g{k}\sigma} \rangle_0\ e^{i(\g i- \g j)\g{k}},\\
  F_{\g{i}\g{j}}&=\frac{1}{ L }\sum_{\g{k}a}\langle \hat f_{\g{k}\sigma}^\dagger\hat f_{\g{k}\sigma} \rangle_0\ e^{i(\g i-\g j)\g{k}},\\
  n_{0f}&=\frac{1}{ L }\sum_{\g{k}a}\langle \hat f_{\g{k}\sigma}^\dagger\hat f_{\g{k}\sigma} \rangle_0.
  \label{lines}
 \end{split}
\end{equation}

To determine the properties of the model, we solve in the self-consistent loop the system of Eqs. (\ref{te}) and (\ref{lines}) 
 \cite{buenemann2012,Jan_hubbard,Jan_tj,PhilMag_Jan,pss} (point 6 from the scheme in Sec. \ref{gsc})

Finally, the ground state energy $E_G$ is defined by
\begin{equation}
 E_G=\langle \mathcal{\hat H}\rangle_G|_0 +n\mu,
\end{equation}
where $\langle \mathcal{\hat H}\rangle_G|_0$ denotes the expectation value (\ref{Ham}) of 
the starting Hamiltonian for the equilibrium values of the 
lines and the total number of particles is defined by 
${n\equiv2\langle \hat n_{\g{i}\sigma}^f+\hat n_{\g{i}\sigma}^c\rangle_G}$.
The $f$-orbital filling separately is defined by ${n_f\equiv2\langle \hat n_{\g{i}\sigma}^f\rangle_G}$.

\section{Results and Discussion}\label{sec3}
\subsection{System description and technical remarks}\label{suba}

 \begin{center}\vspace{-0.1cm}
  \begin{figure}[b]
   \includegraphics[width=0.35\textwidth]{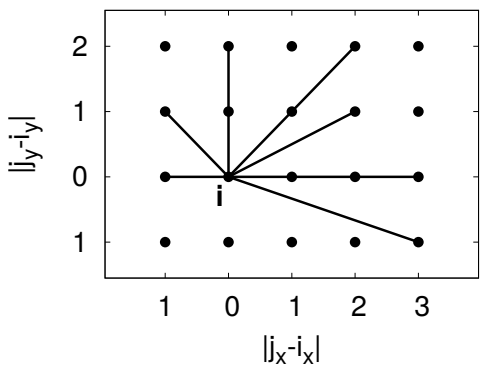}
   \caption{Schematic illustration of the real-space cutoff on the lattice. The solid
   lines denote exemplary, in terms of distance, correlation functions (referred to as {\it lines}) 
   taken into account between $\g i$-site (in the center) and the $\g j$-sites 
   (on the periphery). Farther connections are not considered.}
   \label{Fig2}
  \end{figure}
 \end{center}\vspace{-0.8cm}

In our analysis we consider a square, translationally invariant, and infinite ($ L \rightarrow\infty$) lattice, 
with two orbitals ($f$ and $c$) per site. 
The square lattice consideration is justified by the common quasi-two-dimensional 
layered structure of $f$ atoms in the elementary cell of many 
Ce-based heavy fermion systems \cite{Stewart1984,Pfleiderer2009}
that our studies are relevant to.    

While proceeding with the diagrammatic expansion (DE), in principle two approximations 
need to be made. First, only the lines (\ref{lines1}) satisfying the 
relation $|\g i-\g j|^2 = (i_x-j_x)^2+ (i_y-j_y)^2\leq10$
are taken into account (i.e., we make a real-space cutoff - cf. Fig. \ref{Fig2}). 
For comparison, in VMC only rarely lines farther than these connecting nearest neighboring sites
(more precisely, only the lines corresponding to the hopping term range  
of the starting Hamiltonian) are taken into account \cite{Schmalian,Watanabe}. 
From our numerical calculations it follows that the nearest- and the 
second-nearest neighbor contractions compose the   dominant contributions  (cf. Fig. \ref{Fig6}b).
    
The second limitation in DE is the highest order of the expansion, $k$, taken into account.
Asymptotic behavior starting from $k=3$, of some properties such as the density of states (DOS) 
at the Fermi level (FL), $\rho(E_f)$, 
and the width of the effective $f$ band, $w_f$ (cf. Figs.\ref{Fig3}, \ref{Fig4}c and \ref{Fig5}), 
speak in favor of the calculation reliability, achieved already in 
that order. Therefore, if not specified otherwise, 
the expansion is carried out up to the third order ($k=3$), 
i.e., with the three internal vertices taken into account.
We stress again that the zeroth-order approximation ($k=0$) 
is equivalent to the GA approach (cf. Appendix \ref{B} for details).
The results of GA are regarded here as a reference point for
determining a systematic evolution, including both  
qualitative and quantitative changes, when the higher-order 
contributions  are implemented.

The parameters of the ALM Hamiltonian (\ref{alm})
are taken in units of $|t|$: a strong Coulomb repulsion is taken as $U=10$, 
the reference energy for $f$-electrons, $\epsilon_f=-3$, the onsite hybridization 
is assumed negative and varies in the range $|V|\in(0.8,2.5)$,
and the total band filling (${n\equiv2\langle \hat n_{\g{i}\sigma}^f+\hat n_{\g{i}\sigma}^c\rangle_G}$)
is in the range allowed by the condition that the $f$ level occupancy per site 
($n_{f}\equiv 2\langle\hat n_{\g i \sigma}^f\rangle_G$) 
roughly does not exceed unity. The reason for consideration of this regime  
is the circumstance that for interesting us Ce-based compounds 
the concentration of $f$ electrons per cerium should not exceed 1 (i.e., with the 
Ce$^{3+}$ and Ce$^{4+}$ configurations only).
However, from the construction 
of the method the regime for $n_f>1$ is fully accessible and physically correct.
In carrying out the DE-GWF procedure we adjust the chemical potential $\mu\equiv E_F$
for the fixed total filling $n$.  
Numerical integration of Eq. (\ref{lines})  and the self-consistent loop were 
both performed with precision of the order of $10^{-6}$ or better with the help of 
Gnu Scientific Library (GSL) procedures \cite{GSL}.

 \begin{center}\vspace{-0.1cm}
  \begin{figure} 
   \includegraphics[width=0.45\textwidth]{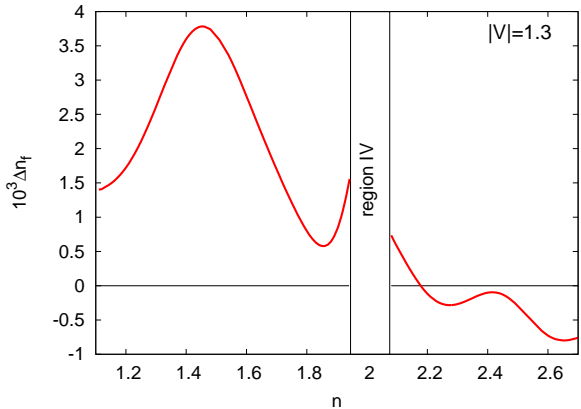}
   \caption{Difference between uncorrelated and correlated $f$-electron number, $\Delta n_f\equiv 
   \sum_\sigma\langle \hat n_{\g i \sigma}^f \rangle_G- \sum_\sigma\langle \hat n_{\g i \sigma}^f \rangle_0$
   along the line of constant hybridization, $|V|=1.3$, with respect to changing total filling.
   The specific character of the region IV is explained in Sec. \ref{sec3}.}
   \label{Fign}
  \end{figure}
 \end{center}\vspace{-0.8cm}

\subsection{Correlated Fermi liquid}
Before the detailed analysis is carried out, a methodological remark is in place. The 
effective Hamiltonian (\ref{effH}) is of single-particle form, but coupled to the self-consistent procedure of
evaluating the relevant averages (\ref{lines}).  However, this does not compose the full picture. 
The physical quantities are those obtained with a projected wave function. For example, $n_f\equiv \sum_\sigma 
\langle\psi_G| \hat f_{\g i\sigma}^\dagger\hat f_{\g i\sigma}|\psi_G\rangle=
\sum_\sigma\langle\mathcal{P}_{G}\hat n_{\g i \sigma}^f\mathcal{P}_{G}\rangle_0$, which
in general is slightly different from $\sum_\sigma\langle \hat n_{\g i \sigma}^f \rangle_0$.
The situation is illustrated explicitly in Fig.~\ref{Fign}. In effect, the quasiparticle picture 
is amended with the nonstandard features of this {\it correlated} ({\it quantum}) {\it liquid} (CL).
Parenthetically, the same difference will appear when considering magnetic and superconducting states,
where the magnetic moments, $\langle\hat S_{\g i}^z\rangle_G$ 
vs. $\langle\hat S_{\g i}^z\rangle_0$, and the superconducting
gaps, $\langle \hat \Delta_{\g i\g j}\rangle_G$ and 
$\langle \hat \Delta_{\g i\g j}\rangle_0$ will be different.
So, we have a mapping of the correlated onto quasiparticle states, 
but not of the physical properties. 
In brief, we have to distinguish between the correlated and the uncorrelated $f$-electron occupancy or other property 
even though, from the way of constructing (\ref{effH}),
the density of quasiparticle states (coming from (\ref{effH})), represents that in the correlated state.

\subsection{Quasiparticle Density of States} 

We start with analysis of the quasiparticle DOS emerging from the 
DE-GWF method  in 
successive orders of the expansion (cf. Fig. \ref{Fig3}). 
For $k>0$ and the total filling $n=1.97$ (i.e., near the half filling),
the hybridization peaks become more pronounced (cf. Fig.\ref{Fig3}-the inset Table) 
and the hybridization gap increases.

\begin{center}\vspace{-0.1cm}
  \begin{figure}[t]
   \includegraphics[width=0.45\textwidth]{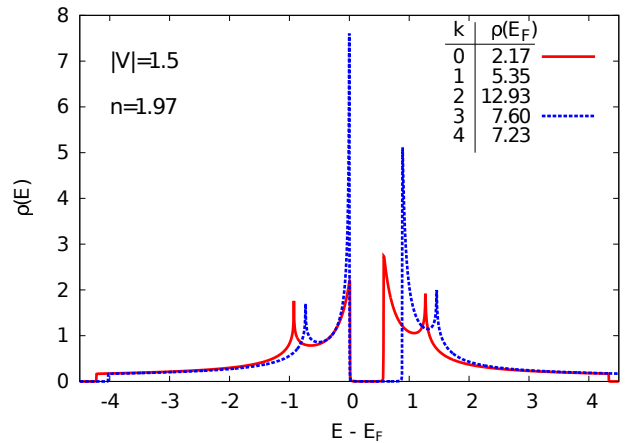}  
   \caption{(Color online) Density of states (DOS) near the half-filling ($n=1.97$) at $|V|=1.5$ 
   for  selected  orders of the diagrammatic expansion ($k=0,3$). 
   Explicit values of $\rho(E_F)$ are also
   listed in the inset Table (for $0\leq k\leq4$). For $k=3$ a 
   satisfactory convergence of the 
   expansion is reached. The $k=1,2,4$ plots are not included
   for clarity as, apart from peak heights, they are practically the same as 
   the plot for $k=3$. 
   For $k>0$ (beyond GA) the hybridization peaks are more pronounced 
   (large DOS at the Fermi level $\rho(E_F)$), which is related directly to 
   the increased by correlation effective-mass enhancement for quasiparticles. 
   }
   \label{Fig3}
  \end{figure}
 \end{center}\vspace{-0.8cm}

For $k>0$ the overall shape of DOS changes only quantitatively (cf. Fig. \ref{Fig3}).
However, the value of the DOS at the Fermi level, $\rho(E_F)$, changes remarkably 
(cf. the inset to Fig. \ref{Fig3}). 
Although for $k=1$ it is underestimated and for $k=2$ overestimated, 
for $k=4$ we see no significant difference with respect to the $k=3$ case. For this reason, if not 
specified explicitly, the subsequent analysis is proceeded in the third 
order, $k=3$.

 \begin{center}\vspace{-0.1cm}
  \begin{figure}[t]
  \vspace{-0.5cm}
  \includegraphics[width=0.5\textwidth]{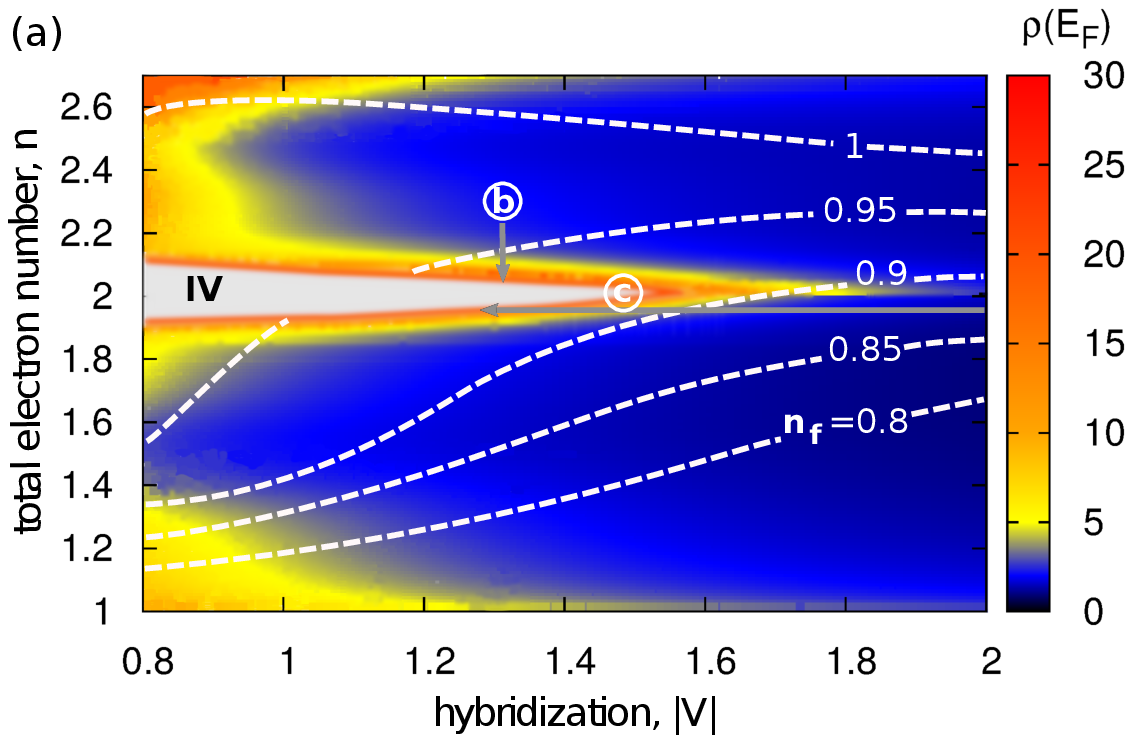}\\
    \includegraphics[width=0.22\textwidth]{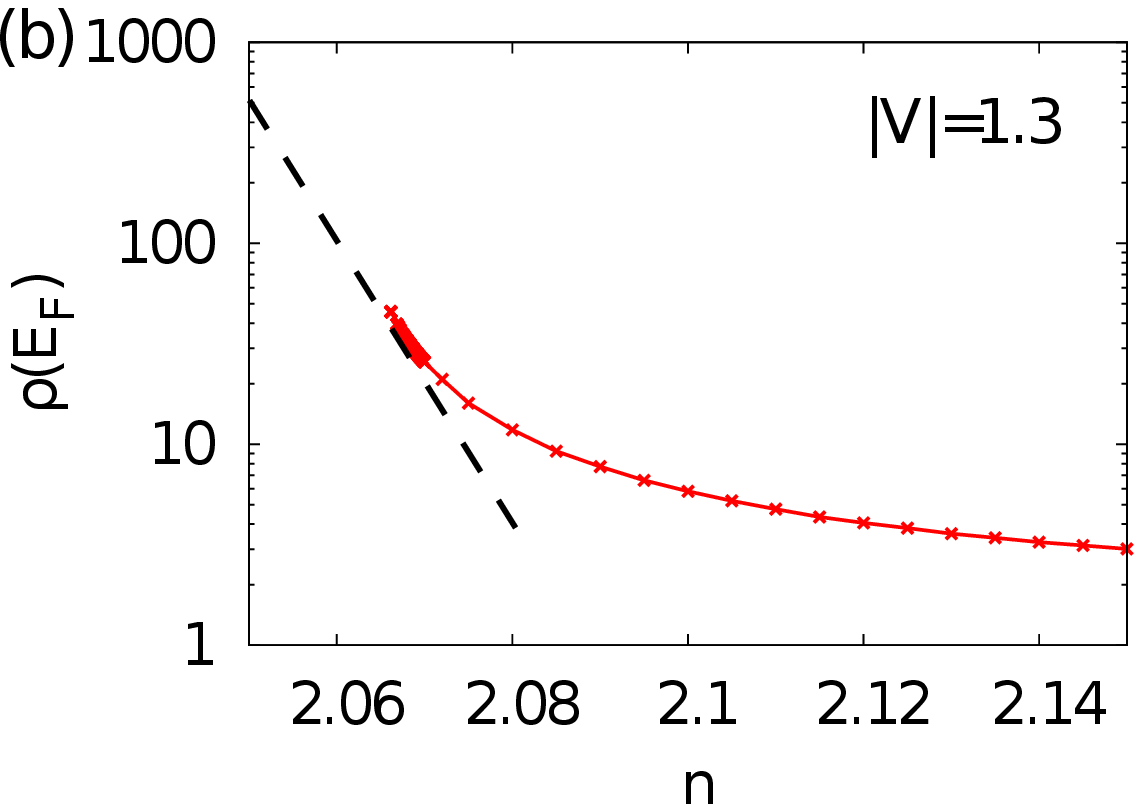}
   \includegraphics[width=0.23\textwidth]{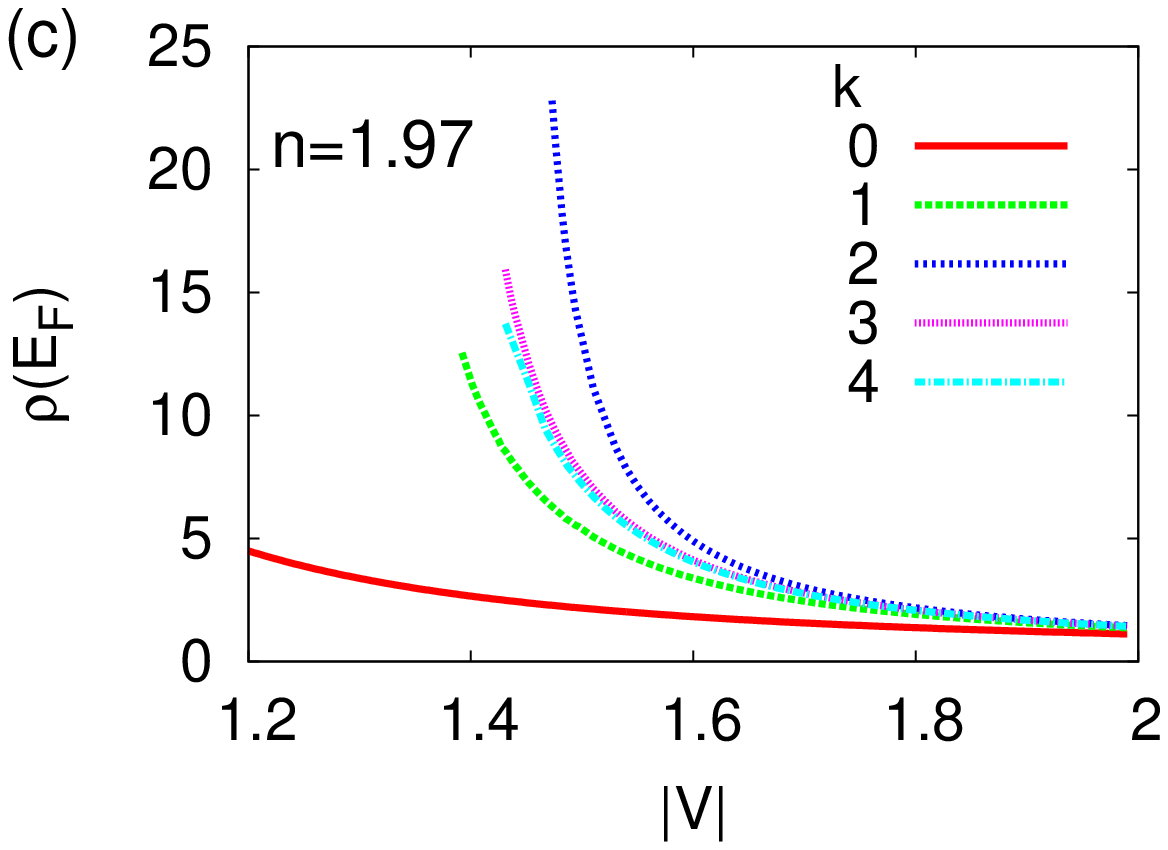}

   \caption{(Color online) (a) Density of states at the Fermi level $\rho(E_F)$
   on the hybridization strength -- total electron concentration plane, $|V|$ -- $n$.
   Additionally (not marked), for $n=2$ we obtain always the Kondo insulating state. 
   By IV (for consistency with Fig.\ref{Fig6}) 
   we have marked a V-shaped region where we have no numerical convergence due to 
   the presence of singular hybridization peaks for low $|V|$ and with $n$ near the half filling  (see main text).
   (b) Evolution of $\rho(E_F)$ in the half-logarithmic scale near the region IV
   (along the vertical arrow with the letter b).
   By extrapolation (dashed line in (b)), for the almost half-filled situation, $\rho(E_F)$ can 
   be enhanced even by factor of 1000 relative to its lowest values on the 
   $|V|$ -- $n$ plane. (c) Evolution of $\rho(E_F)$ 
   with the decreasing $|V|$ (along the horizontal arrow with letter c),
   within successive orders of the expansion ($k\leq4$). For large $|V|\gtrsim1.8$, GA ($k=0$ order) 
   provides already realistic values of $\rho(E_F)$.}
   \label{Fig4}
  \end{figure}
 \end{center}\vspace{-0.8cm}

The value of $\rho(E_F)$ is of crucial importance. 
This parameter is a measure of the quasiparticle 
effective mass, as the latter is inversely proportional 
to the second derivative of the energy, 
$\nabla^2_{\g k}E_{\g k}$, at the Fermi surface, and thus is 
determined by $\rho(E_F)$. 

In Fig. \ref{Fig4}a we draw the value of $\rho(E_F)$ on the plane 
hybridization -- total electron number (per site), $V$ -- $n$. 
This quantity is particularly strongly enhanced near the half filling ($n\simeq 2$).
In comparison to the lowest value $\rho(E_F)\approx 0.75$, the maximal enhancement  is of the order of 40.
In Fig. \ref{Fig4}b we present evolution of $\rho(E_F)$ on the logarithmic scale
with the decreasing total filling and approaching $n=2$ (vertical arrow in   Fig.~\ref{Fig4}a 
marked  by the encircled letter b ). The extrapolated value of
$\rho(E_F)$ may reach extremely high values of 1000 and even more 
(dashed line in Fig. \ref{Fig4}b)
in the region IV.
Such feature could explain extremely high 
mass renormalization in some of HFS for
large but finite value of the Coulomb interaction $U$.

The region where $\rho(E_F)$ is enhanced strongly, is that with low hybridization
$|V|$ values and for the total filling $n\simeq 1$. This region is strictly correlated with
 the position of the second pronounced peak in DOS (cf. Fig. \ref{Fig3}) 
 which therefore has its meaning as the Van Hove singularity.
 Additionally, for $n_{f}\simeq 1$, where the effects of correlations are the strongest, 
 we observe also a large value of $\rho(E_F)$. In that limit the stability of magnetic 
 phases should be studied separately
 \cite{Howczak2012,Howczak2013}.
 
As marked in Fig. \ref{Fig4}a, near the total half-filling, $n\simeq2$, 
we could not obtain a satisfactory convergence of our self-consistent procedure.
This is attributed to the position of the chemical potential
extremely close to the hybridization-induced peaks (significant when $n_{f}\gtrsim 0.9$).
Technically, this leads to extreme fluctuations (out of our numerical precision) 
of the filling, effective hopping parameters, 
and the lines coming from the effective Hamiltonian (\ref{effH}), as they 
are sensitive to a slight change of the chemical potential position.
For $n=2$ and nonzero hybridization, we obtain always the Kondo insulating state.
However, strictly speaking, the true Kondo-type compensated state is demonstrated explicitly
only if magnetic structure is taken into account explicitly \cite{Doradzinski1997,Doradzinski1998,Howczak2012}.

In Fig.~\ref{Fig4}c we depict the $\rho(E_F)$ evolution with the decreasing 
hybridization amplitude $|V|$ for $k\leq4$. Our results show that for large $|V|$, 
GA ($k=0$) already is reasonable approximation. The situation changes 
as we approach the low-$|V|$ regime near the half-filling, where inclusion
of higher-order contributions leads to a strong enhancement of  $\rho(E_F)$, as discussed above.   

 In summary, the quasiparticle mass is enhanced spectacularly near $n=2$ and in the regime of small hybridization $|V|$.
 The $f$-state occupancy is then $n_f\gtrsim0.9$. This is the regime associated with the heavy-fermion 
 and the Kondo-insulating states. We discuss those states in detail in what follows.
 
 \subsection{$f$-electron direct itineracy}
  \begin{center}\vspace{-0.1cm}
  \begin{figure}[t]
   \includegraphics[width=0.45\textwidth]{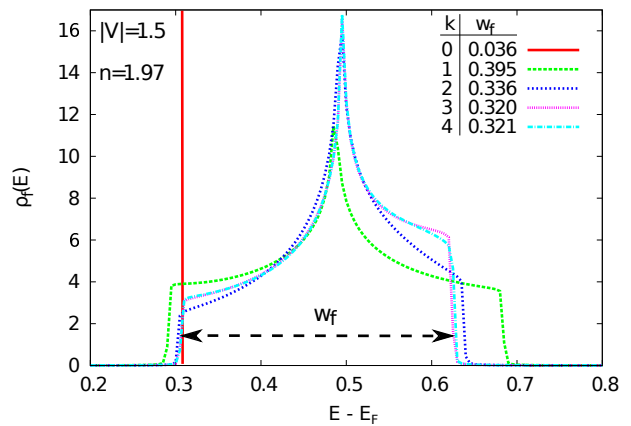}
   \caption{(Color online) $f$-electron density of states $\rho_f(E)$ 
   within successive orders of expansion ($k\leq4$). 
   For $k=1$ and higher, formation of the effective $f$-band can be clearly observed.
   For  $k=3$ the final shape of $\rho_f(E)$ and the value of   $f$-band 
   width $w_f$ stabilize.}
   \label{Fig5}
  \end{figure}
 \end{center}\vspace{-0.8cm}
 
As stated already, the DE-GWF method is used here to map the correlated (many-body) system, described by
 the original Hamiltonian (\ref{alm}) with the help of the Gutzwiller wave function $|\psi_G\rangle$,
 onto that described by the effective quasiparticle Hamiltonian (\ref{effH}) with an
 uncorrelated wave function $|\psi_0\rangle$. 
 By  constructing the 
 effective Hamiltonian it is possible to extract the explicit contribution
to the quasiparticle picture as coming from a direct hopping between the neighboring $f$ sites. 
 By contrast, in GA ($k=0$) case, the $f$ electrons itineracy is only due to the admixture of $c$-states when the quasiparticle
 states are formed. 
Once we proceed with the diagrammatic expansion to higher order ($k>0$), they 
 start contributing to the quasiparticle spectrum in the form of a dispersive $f$-band (cf. Fig. \ref{Fig5}).
 The resulting band is narrow, $w_f\leq0.5$, whereas the starting conduction ($c$) band has the width of $w_c=8$.
As was mentioned in the Sec. \ref{I}, we interpret the
 parameter $w_f$ as a measure of emerging degree of {\it direct itineracy}, i.e., 
presence of a direct hoppings between the neighboring $f$ states
 in the effective Hamiltonian.
 
 Again, a methodological remark is in place here on the numerical convergence of the results with respect to $k$.
 Namely, the $f$-bandwidth appears already for $k=1$, but both its width and the 
 curvature stabilizes only starting from $k=3$.
  
  In the recent phenomenological modeling of CeCoIn$_5$ \cite{Yazdani2012,allan2013,dyke2014} 
  the band structure used is the hybridized-two-band independent-particle model 
  with dispersive $f$-band, even though
  the Ce 4$f$ states can be placed well above the so-called Hill limit, where there should not be any 
  direct hopping between the original neighboring $f$ states. 
  The fit presented there provides $w_f$ of the same order of magnitude 
  as that obtained here. 
  As those phenomenological models do not include the Coloumb interaction, the ground state  is determined 
  by the uncorrelated wave function. Hence, our analysis of the effective Hamiltonian
  resulting from ALM provides a direct microscopic rationalization of  
  the narrow dispersive $f$-band presence assumed {\it ad-hoc} 
  in the fitting procedure in Ref.~\onlinecite{Yazdani2012,allan2013,dyke2014}.

    \begin{center}\vspace{-0.1cm}
  \begin{figure}[t]
  \vspace{-0.5cm}
      \includegraphics[width=0.5\textwidth]{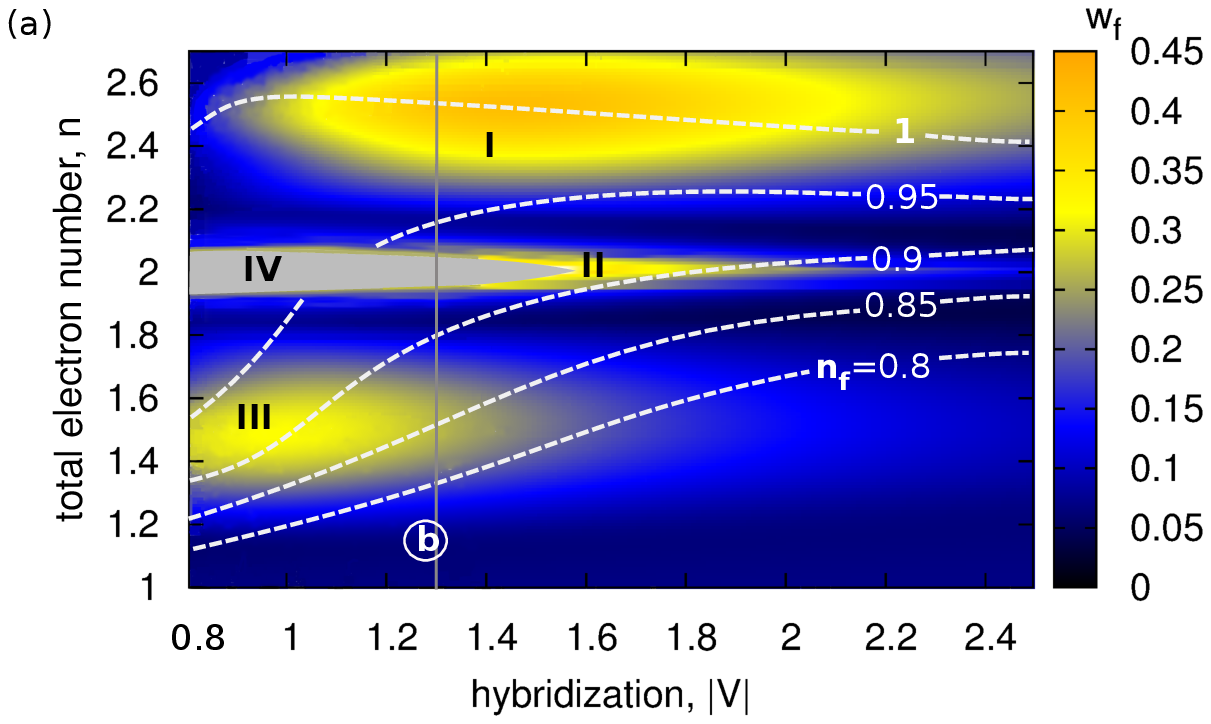}  
      \includegraphics[width=0.45\textwidth]{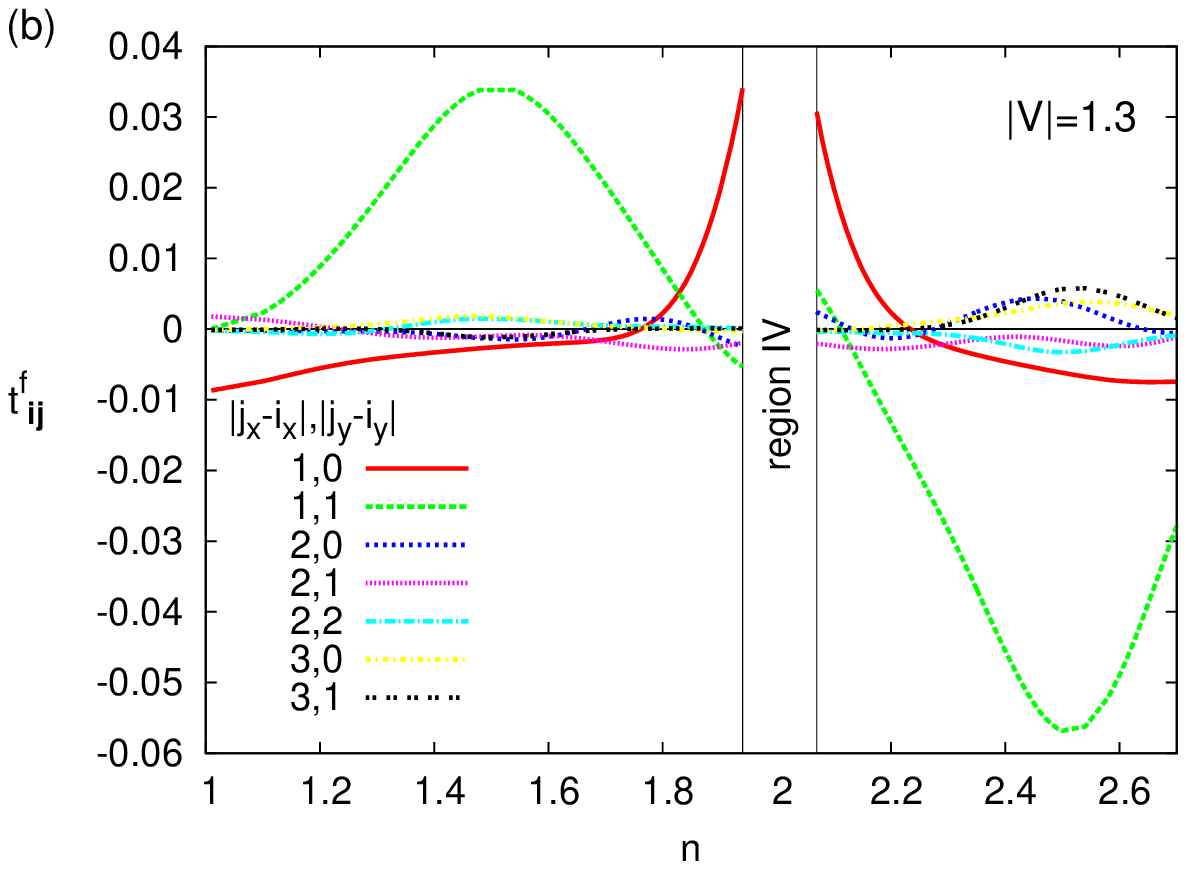}
    \caption{(Color online) (a) Effective bandwidth of $f$-states, $w_f$, on 
    the hybridization strength $|V|$ -- 
   electron concentration $n$  plane. $w_f$ is regarded as a 
   measure of direct itineracy of $f$-electron states. 
   Three separate disjoint regions (light color) are regarded as universal and 
   frequently discussed as separate limits, both 
   in theory and experiment. Namely, the 
   mixed valence regime (III), the Kondo/almost Kondo-insulating regime (II), and 
   the Kondo-lattice regime (I) with $n_{f}\rightarrow1-\delta$, $\delta \ll 1$.
   (b) Effective $f$-electron intersite hoppings $t^f_{\g i\g j}$ along the marked 
   vertical line of the diagram for $|V|=1.3$.
   The energy dispersion for $f$ quasiparticles is determined mainly by the 
   nearest and the second nearest hoppings $t^f_{\g i\g j}$. Region IV, near $n=2$ is marked 
   separately due to the lack of convergence of the numerical results (see main text).
    }
   \label{Fig6}
  \end{figure}
 \end{center}\vspace{-0.8cm}
 
 In Fig.~\ref{Fig6}a we display diagram comprising the width of 
 $f$-band $w_f$ on $|V|$ -- $n$ plane, with contours of constant values 
 of $n_{f}$. We observe the appearance of regions, where the $f$ quasiparticles have 
 a sizable bandwidth (bright color) and other, where they remain localized (dark regimes).
 We expect that in the regions, where $f$ electrons are forming a band,
a nontrivial unconventional superconductivity and/or magnetism may appear. 
These topics should be treated separately as they require a substantial extension
of the present approach (incorporating new type of lines) \cite{Jan_hubbard,Jan_tj,PhilMag_Jan,pss}.
 
 With the help of the width $w_f$ we may single out three physically 
 distinct regimes (cf. Fig. \ref{Fig6}a).
 We identify those regions as the mixed-valence regime (III), the Kondo/almost Kondo insulating regime (II),
 and the Kondo-lattice regime (I) with $n_{f}\rightarrow1-\delta$, with $\delta\ll1$ 
 (cf. Fig.~\ref{Fig6}a). These universal regions
 are usually discussed independently within different specific models and methods.
 In regime I the role of $f$-$c$ Coulomb interactions (the Falicov-Kimball term)
 may be needed for completeness (cf. Ref. \onlinecite{Miyake2000}), whereas in the Kondo-lattice 
 regime the transformation 
 to the Anderson-Kondo model is appropriate (cf. Refs. \onlinecite{Howczak2012,Howczak2013}).
In the extreme situation, the heavy-fermion states are modeled by pure Kondo-lattice model\cite{Spalek2015, Doniach1987,
Lacroix2003}. However,
strictly speaking, the last model applies only in the limit of localized $f$ electrons ($n_f=1$), 
since then the total numbers of $f$ and $c$ electrons are conserved separately.
 
 In Fig. \ref{Fig6}b we present the
 effective hopping parameters for $f$ states for $|V|=1.3$, i.e., along the 
 marked vertical line in Fig. \ref{Fig6}a.
 This line crosses three singled out regions of the itineracy.
 The leading contribution to the $f$-electron band energy arises from
 the nearest- and the second nearest-neighbor hoppings. Such circumstance confirms 
 that our  earlier assumption about the real-space cutoff shown in Fig. \ref{Fig2} 
 has been selected properly. 
 Moreover, it points to the importance of including also the  
components beyond those of the starting Hamiltonian, 
only rarely taken into account within the VMC method \cite{Watanabe,Schmalian}.

 In Fig. \ref{Fig_v} we show the contributions to the effective hybridization.
The initial (bare) local hybridization acquires momentum dependence. Nevertheless, the 
 local part is still dominant since the nonlocal terms are at least two 
 orders of magnitude smaller.
 
   \begin{center}\vspace{-0.1cm}
  \begin{figure}[t]
   \includegraphics[width=0.45\textwidth]{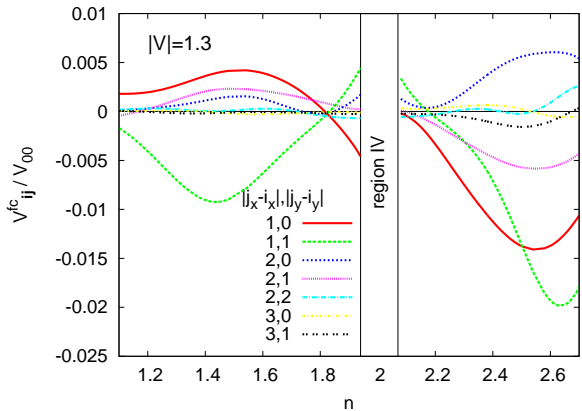}
   \caption{(Color online) Spatial contributions $V^{cf}_{\g i\g j}$ to the effective hybridization normalized by the first
   onsite ($\g i=\g j$) $V_{00}$ term along the marked 
   vertical line of the diagram in Fig.\ref{Fig6}a for $|V|=1.3$.
   Note that due to correlation initially local, onsite hybridization acquires effectively
   momentum dependence. However, the nonlocal contributions constitute only up to 2\% of 
   the local one.}
   \label{Fig_v}
  \end{figure}
 \end{center}\vspace{-0.8cm}

 The emerging in our model $f$-band introduces a new 
 definition of the $f$-electron itineracy as it is not so much connected
 to the Fermi-surface size \cite{Hoshino2013}, but with the
 appearance of a direct hoppings between $f$ sites. This difference is highly nontrivial, especially 
in the limit $n_{f}=1$, where we obtain the largest  bandwidth $w_f$.
Such behavior is attributed to the specific character of our approach. Namely,
we consider here the processes within our initial Hamiltonian (\ref{alm}),
but under the assumption that the neighboring sites are also correlated. This, as we have shown directly, 
leads also to the finite $f$-band in the effective single particle Hamiltonian (\ref{effH}).
The results thus throw a new light on the longstanding issue of the dual localized-itinerant nature of $f$ electrons 
in HFS \cite{Park2008,Troc2012}. While the
magnetism can be attributed to the almost localized nature of $f$ electrons, an unconventional superconductivity requires their
itineracy in an explicit manner, as will be discussed elsewhere \cite{sc_alm}  
 
\section{Summary}
We have applied a recently developed diagrammatic technique (DE-GWF) of   
evaluating  the expectation values with the full Gutzwiller wave function
for the case of two-dimensional Anderson lattice. 
We have analyzed
properties of the model by discussing the most important features of the 
heavy fermion systems in the pa\-ra\-mag\-ne\-tic state. 
We have also shown that by approaching in successive orders of the expansion
the full Gutzwiller-wave-function 
solution, we obtain a systematic convergence. 
In the zeroth order of expansion our method 
reduces to the standard Gutzwiller Approximation (GA).

In difference with GA, DE-GWF does not overestimate 
 the hybridization narrowing factor. Furthermore, our method
produces unusually enhanced peaks at the Fermi 
level in the density of states, particularly near the half-filling, $n\rightarrow2$. 
This in turn, is connected to the value of effective mass and by analyzing in detail this 
region we can explain a very large mass enhancement observed in heavy fermion systems as
described by ALM with large, but finite Coulomb-interaction value, here $U=10|t|$. 
The regions of sizable $\rho(E_F)$ enhancement are also found in the small-hybridization limit 
and are connected to the presence of both the Van Hove singularity and the
strong correlations in the limit of $n_{f}\rightarrow 1$.

The $f$-electron contribution to the full quasiparticle spectrum is analyzed in detail.
For nonzero order of the expansion ($k>0$) we observe a systematic formation of the effective $f$-band
with the increasing $k$.   
In spite of the fact that the bare electrons are  initially localized, $f$ 
quasiparticles contribute to the total density of states
as they become itinerant. We interpret this property as the emerging {\it direct}
$f$-electron {\it itineracy}. As a measure of this behavior, we introduce the 
the width  $w_f$ of effective $f$-band. 
Formation of such narrow $f$-band rationalizes e.g.
the recent phenomenological modeling of the CeCoIn$_5$ band structure 
\cite{Yazdani2012,allan2013,dyke2014}.

The nonstandard character of the resultant {\it correlated} Fermi {\it liquid} (CL) which differs 
from either the Landau {\it Fermi liquid} (FL) and the {\it spin liquid} (SL), should be stressed. FL represents 
a weakly correlated state (no localization) and SL represents a fully correlated state. Our 
CL state in this respect has an intermediate
character. Namely, the quasiparticle states are formed (as exemplified by e.g. density of states),
but the physical properties such as the occupancy $n_f$, the magnetic moment $\langle \hat S^z_{\g i}\rangle$
or the pairing gap in real space $\langle \hat \Delta_{\g i\g j}\rangle$ are strongly renormalized by the correlations.
Such situation is often termed as that of an almost localized Fermi-liquid state 
\cite{Fulde1988,Doradzinski1997,Doradzinski1998,Hewson1993,Auerbach1987}. 

By analyzing the results on the hybridization strength $|V|$ -- total band filling  $n$ plane,
we single out explicitly three physically distinct regions, which we regard as three separate
universality limits.
Namely, we have linked those disjoint regions with the regimes frequently 
discussed as separate classes in the heavy fermion systems: the mixed-valence regime, 
the Kondo/almost Kondo insulating regime, and the Kondo-lattice regime for $n_{f}\rightarrow1$. 
We suggest,  that the regions of significant $f$-electron itineracy can be connected to the
unconventional heavy fermion superconductivity which would require separate studies.

We have also commented on the longstanding issue of a dual localized-itinerant
nature of $f$ electrons in the heavy fermion
systems. The new definition of itineracy 
is in accord with their (almost) localized nature.  

\section*{Acknowledgements}
We are grateful for discussions with  J\"org B\"unemann.
The work was partly supported by the  
National Science Centre (NCN) under the Grant MAESTRO, No. DEC-2012/04/A/ST3/00342.
Access to the supercomputer located at Academic Center for 
Materials and Nanotechnology of the AGH University of Science and Technology in Krak\'ow is also acknowledged.
MW acknowledges also the hospitality of the Institute of Science and Technology Austria during the 
final stage of development of the present work, as well as a partial 
financial support from Society - Environment - Technology project
of the Jagiellonian University for that stay. 
JK acknowledges support from the People Programme (Marie Curie Actions) of the 
European Union's Seventh Framework Programme (FP7/2007-2013) under REA grant agreement n$^{\rm o}$ [291734].
 
\appendix
\section{Equivalence of the k=0 order DE-GWF expansion and the Gutzwiller approximation (GA)}\label{B}
Here we show the equivalence of the zeroth order DE-GWF and the standard Gutzwiller approximation (GA).
In both methods (DE-GWF in the zeroth order of expansion $k=0$) the effect of the projection 
can be summarized by the expressions for evaluating following expectation values: 
$\langle\hat n_{\g i \uparrow}\hat n_{\g i \downarrow}\rangle_G$ and 
$\langle  \hat f^\dagger_{\g i\sigma}\hat c_{\g i\sigma}+\text{H.c.} \rangle_G$.
The remaining averages in ALM are unchanged under the projection.

Explicitly, in the DE-GWF for $k=0$ the resulting averages are expressed as follows  
\begin{subequations}
\begin{align}
 \langle\hat n_{\g i \uparrow}\hat n_{\g i \downarrow}\rangle_G^{(k=0)}
 &= \lambda_d^2n_{0f}^2 \label{B3a}\\
  \langle  \hat f^\dagger_{\g i\sigma}\hat c_{\g i\sigma}+\text{H.c.} \rangle_G^{(k=0)}
 &=  \alpha \langle \hat f^\dagger_{\g i\sigma} \hat c_{\g i\sigma} +\text{H.c.}\rangle_0,
 \end{align}
\end{subequations}
where parameter $\alpha$ (see also Appendix~\ref{A}: Eqs. (\ref{A11}) and (\ref{A4})) is defined as 
\begin{equation}
 \alpha \equiv (1-n_{0f})\lambda_0\lambda_s+n_{0f}\lambda_d\lambda_s.\label{al}
\end{equation}

On the other hand, in GA the resulting averages are expressed as\cite{Rice1985}
\begin{subequations}
\begin{align}
 \langle\hat n_{\g i \uparrow}\hat n_{\g i \downarrow}\rangle_G^{(GA)}
 &= \langle n^f_{\g i \uparrow}\hat n^f_{\g i \downarrow}\rangle_0\equiv  d^2,\label{B1a} \\
  \langle  \hat f^\dagger_{\g i\sigma}\hat c_{\g i\sigma}+\text{H.c.} \rangle_G^{(GA)}
 &= \sqrt{q} \langle \hat f^\dagger_{\g i\sigma} \hat c_{\g i\sigma} +\text{H.c.}\rangle_0,\label{B1b} 
 \end{align}
\end{subequations}
where the parameter $d^2$ is the double occupancy probability, 
and $q$ is the so-called Gutzwiller factor reducing the hybridization amplitude, which
for the equal number of particles for each spin   
is defined as
\begin{equation}
 \sqrt{q}=\frac{\sqrt{(n_{0f}-d^2)(1-2n_{0f}+d^2)}+\sqrt{(n_{0f}-d^2)d^2}}{\sqrt{n_{0f}(1-n_{0f})}}.\label{B2}
\end{equation}

If we identify double occupancy probabilities expressed by both methods in (\ref{B1a}) and (\ref{B3a}) 
to be equal, yielding $d^2=\lambda_d^2n_{0f}^2$, then the parameter $\alpha$ (\ref{al}) exactly 
reduces to the parameter $\sqrt{q}$ (\ref{B2}).

GA procedure results in the effective single-particle Hamiltonian of the form
\begin{equation}
\begin{gathered}
 \mathcal{\hat H}_{GA} \equiv  \sum_{\g{k},\sigma}  \hat\Psi^\dagger_{\g{k}\sigma}
\begin{pmatrix}
 \epsilon_{\g{k}}^{c} -\mu&\sqrt{q_\sigma}V \vspace{3pt}\\
 \sqrt{q_\sigma}V& \epsilon_{f}-\mu   \\ 
\end{pmatrix}
\hat\Psi_{\g{k}\sigma}
+ L  Ud^2\\  - \lambda^f_n \Big( \sum_{\g{k},\sigma}\hat n^f_{\g{k},\sigma}-  L  n_{0f} \Big)
 -\lambda^f_m \Big( \sum_{\g{k},\sigma}\sigma\hat n^f_{\g{k},\sigma} -   L  m_f \Big) .
\label{HSGA}
\end{gathered}
\end{equation}
In the above Hamiltonian it is necessary to add constraints for $f$-electron concentration and their magnetization 
in order to satisfy consistency of the procedure \cite{sga,Rice1986}.  
In effect, the whole variational problem is reduced to minimization of the ground state energy with respect to $d^2$, $n_{0f}$, $m_f$,
and respective Lagrange multipliers $\lambda_n^f$ and $\lambda_m^f$, playing  
the role of the effective molecular fields \cite{sga}. However, the effect of constraint for $f$-electron
magnetization is relevant only in the case of magnetism 
consideration either as intrinsic\cite{WysokinskiAbram2014, WysokinskiAbram2015} or
induced by applied magnetic field \cite{Wysokinski2014}. 
Here, as we discuss paramagnetic state $m^f=\lambda_m^f=\lambda_n^f=0$. 

The DE-GWF method by construction guarantees
that the variationally obtained $f$-electron occupancy number $n_f$ 
coincides with that obtained self-consistently\cite{PhilMag_Jan}. 
We have thus provided analytical argument for the equivalence of the DE-GWF method for $k=0$
and the standard GA procedure. Also, by an independent numerical crosscheck we have verified
that all the observables calculated within both methods indeed coincide.

 \section{Diagrammatic sums}\label{A}
  \begin{center} 
  \begin{figure}[t]
   \includegraphics[width=0.45\textwidth]{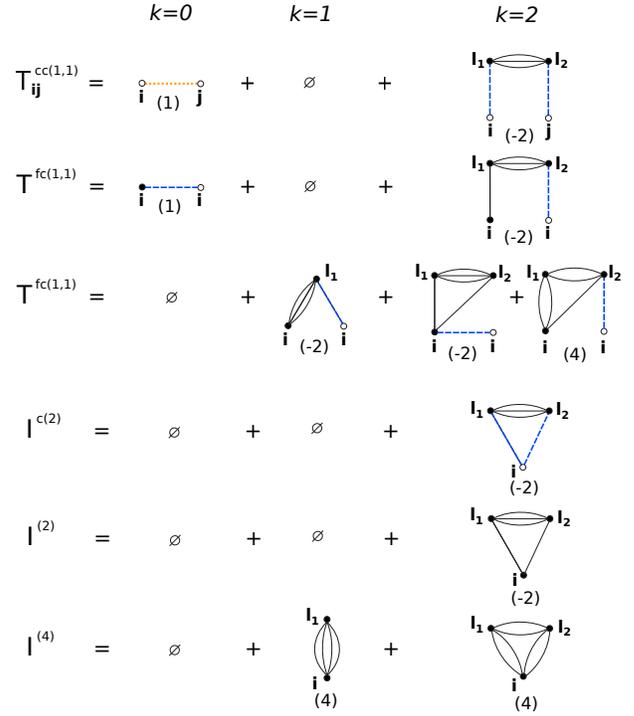}
   \caption{(Color online) Diagrammatic sums to the second order, $k=2$. 
   $c$- and $f$- orbital sites are denoted with empty and filled 
   circles respectively. Solid, dashed (blue) and dotted (orange) 
   connections  represent  $F$, $W$, and $C$ lines respectively (cf. Eq. (\ref{lines1})).
   The numbers in brackets under diagrams stand  for their multiplicity resulting from the Wick's theorem. Note that by construction
   of our sums we have no diagrams with so-called ``Hartree bubbles'', namely loop-lines within the same site and orbital. }
   \label{Fig7}
  \end{figure}
 \end{center} 
We start with expressions for the following projected operators originating from ALM Hamiltonian (\ref{alm}), namely

\begin{equation}
\begin{split}
\hat P_{G;\g i} \hat d_{\g i} \hat P_{G;\g i}&=\lambda_d^2[2n_{0f}\hat n^{HF}_{\g i}+(1-xd_0)\hat d_{\g i}^{HF}+d_0\hat P_{G;\g i}^2], \vspace{2pt} \\
\hat P_{G;\g i} \hat n_{\g i\sigma} \hat P_{G;\g i}&=(1+xm)\hat n^{HF}_{\g i}+\gamma\hat d_{\g i}^{HF}+n_{0f}\hat P_{G;\g i}^2,  \vspace{2pt} \\
\hat P_{G;\g i} \hat f_{\g i\sigma}^{(\dagger)} \hat P_{G;\g i}&=\alpha\hat f_{\g i\sigma}^{(\dagger)}+
\beta\hat f_{\g i\sigma}^{(\dagger)}\hat n^{HF}_{\g i}, \label{A11}
\end{split}
\end{equation}
where  additionally we have defined
\begin{equation}
\begin{split}
\hat n^{HF}_{\g i}&\equiv\hat n^{HF}_{\g i\sigma}=\hat n^{HF}_{\g i\bar\sigma},\\
\beta&\equiv\lambda_{s}( \lambda_d-\lambda_0),\\
\alpha&\equiv\lambda_s \lambda_0+\beta n_{0f},\\
\gamma&\equiv x(1-2n_{0f}),\\
d_0&\equiv n_{0f}^2,\\
m&\equiv n_{0f}(1-n_{0f}).\label{A4}
\end{split}
\end{equation}
As mentioned in the main text, such form of the projected operators significantly speeds up the 
convergence of the numerical results \cite{buenemann2012}, since by construction all two-operator 
averages for a single site and $f$-orbital, the so-called {\it Hartree bubbles}, vanish.   
The above operator algebra leads to the compact definition of the diagrammatic sums:  
 ${S\in\{T^{cc(1,1)}_{\g{ij}}, T^{fc(1,1)}_{\g{ij}}, T^{fc(3,1)}_{\g{ij}},I^{c(2)}_{}, I^{(2)}_{},I^{(4)}\}}$
in Eq.~(\ref{Ham}),
\begin{equation}
 S=\sum_{k=0}^\infty\frac{x^k}{k!}S(k).
\end{equation}
with the $k$-th order contributions
\begin{equation}
\begin{split} 
  T^{cc(1,1)}_{\g{ij}}(k)&\equiv\sum_{\g l_1,...,\g l_k}\langle\hat c^\dagger_{\g i\sigma}\hat c_{\g j\sigma}\hat d^{HF}_{\g l_1,...,\g l_k}\rangle_0^c,\\
 T^{fc(1[3],1)}(k)&\equiv\sum_{\g l_1,...,\g l_k}\langle[\hat n^{HF}_{\g i}]\hat f^\dagger_{\g i\sigma}\hat c_{\g i\sigma}\hat d^{HF}_{\g l_1,...,\g l_k}\rangle_0^c,\\
I^{c(2)}(k)&\equiv\sum_{\g l_1,...,\g l_k}\langle\hat n^c_{\g i\sigma}\hat d^{HF}_{\g l_1,...,\g l_k}\rangle_0^c,\\
  I^{(2)}(k)&\equiv\sum_{\g l_1,...,\g l_k}\langle\hat n^{HF}_{\g i}\hat d^{HF}_{\g l_1,...,\g l_k}\rangle_0^c,\\
  I^{(4)}(k)&\equiv\sum_{\g l_1,...,\g l_k}\langle\hat d^{HF}_{\g i}\hat d^{HF}_{\g l_1,...,\g l_k}\rangle_0^c.\label{a48}
\end{split}
\end{equation}
Superscript $c$ in the expectation values means that only the connected diagrams are to be included.
Note that in (\ref{a48}) there are no summation restrictions, due to the linked cluster theorem \cite{Fetter}.
 The resulting diagrammatic sums for $S$ up to second order ($k=2$) are depicted in Fig. \ref{Fig7}.
 

\begin{thebibliography}{83}
\bibitem[{not()}]{nota}
\bibinfo{note}{In principle, the hybridization can have an intersite character
  \cite{Coleman}.}
\end{thebibliography}

\end{document}